Title

# Gravitational disturbance on asteroidal ring systems by close encounter with a small object

Authors

**Ren Ikeya [a*] and Naoyuki Hirata [a]**

\* Corresponding Author E-mail address: r-ikeya@stu.kobe-u.ac.jp

**Authors' affiliation**

[a] Graduate School of Science, Kobe University, Kobe, Japan.

**Proposed Running Head: Asteroidal ring under close encounter**

Editorial Correspondence to:
Mr. Ren Ikeya
Kobe University, Rokkodai 1-1 657-0013
Tel/Fax +81-7-8803-6566

**Highlights**

- Rings are undisturbed unless objects with comparable or greater mass pass closely.
- Disruptive close encounters are unlikely to occur in 4 Gyr.
- Ring systems around larger objects have a longer lifetime.

**Abstract**

To date, rings are found around a Centaur (10199) Chariklo, trans-neptunian objects (TNOs) (136108) Haumea, and (50000) Quaoar. These discoveries suggest that asteroidal ring systems may be common, particularly in the outer solar system. Since collisions are a ubiquitous and fundamental evolutionary process throughout the solar system, we conjecture that asteroidal ring systems must have experienced close encounters with small objects as part of their evolutionary process. Here, we investigate the response of ring systems when they experience gravitational disturbance by a close encounter with another small object, by calculating the change in eccentricity and the fraction of lost ring particles. We find that a perturber needs to be as massive as or more massive than the ringed object, and needs to pass in the immediate vicinity of the ring in order to cause significant disruption. The change in eccentricity expected for Chariklo's inner ring and Quaoar's outer ring agrees with the analytical expression derived from the impulse approximation, while that for Haumea's ring agrees with the analytical expression called "exponential regime". If we define a lifetime of a ring as the mean time to experience disruptive close encounters that can raise the eccentricity of ring particles greater than 0.1, the lifetime for ring systems around Chariklo, Haumea, and Quaoar are $>10^4$ Gyr. We conclude that ring systems around Chariklo, Haumea, and Quaoar are highly unlikely to suffer from close encounters with another small object even if those systems are as old as 4 Gyr. Close encounter with a small object may not be responsible for the finite eccentricity observed for Chariklo's ring system, although eccentricity of ring particles could be increased to $\sim 10^{-4}$ in 4 Gyr. We also find that the lifetime is shorter for smaller ringed objects, and it still exceeds 4 Gyr for km-sized ringed objects in the outer solar system. Therefore, regardless of the size of ringed objects, asteroidal ring systems in the outer solar system are unlikely to suffer severe damage by close encounter with a small object. On the other hand, the lifetime is estimated to be on the order of 100 Myr for km-sized asteroids and 10 Myr for 100 m-sized asteroids in the near-Earth region and the main belt. This finding offers a dynamical explanation why ring systems have not been found in the inner solar system.

## 1. Introduction

Asteroidal ring systems are found around Chariklo (Braga-Ribas et al., 2014), Haumea (Ortiz et al., 2017), and Quaoar (Morgado et al., 2023; Pereira et al., 2023). Small objects in the outer solar system, such as Centaurs and TNOs, are expected to be remnants of the early solar system and therefore expected to preserve its evolutionary history. Studying formation mechanisms and stability of asteroidal ring systems should shed light on the evolution of both small objects themselves and the solar system. Furthermore, studies on asteroidal ring systems can provide insights into the origin and evolution of planetary ring systems since they share similar dynamics.

The first detection of the asteroidal ring system was around Chariklo, whose ring system consists of two narrow dense rings (Braga-Ribas et al., 2014). Variation in the width was observed and thus the inner ring is inferred to have a finite eccentricity (Braga-Ribas et al. 2014; Pan & Wu, 2016) like the uranian rings (e.g., Goldreich & Tremaine, 1979a, 1979b). The origin of eccentricity is still unclear. Centaur 95P/(2060) Chiron was proposed to have a ring system to explain the previously acquired occultation data (Ortiz et al., 2015). However, the presence of its ring system is currently under debate because Chiron's cometary activity may be responsible (Luu

& Jewitt, 1990; Elliot et al., 1995; Bus et al., 1996; Ruprecht, 2013; Ruprecht et al, 2015). Haumea was also found to possess a narrow and dense ring (Ortiz et al., 2017). The ring lies close to the 3:1 mean motion resonance with Haumea's spin rate, which means that the ring particles complete their orbit once while Haumea spins three times. Recent occultations revealed that Quaoar has two rings: the inner ring (Pereira et al., 2023) and the dense and inhomogeneous outer ring (Morgado et al., 2023). Both rings reside outside the classical Roche limit. Morgado et al. (2023) conducted local *N*-body simulations with a soft-particle method (Salo, 1995) and showed that ring particles can avoid gravitational accretion even outside the classical Roche limit, assuming a velocity-dependent restitution coefficient based on laboratory experiments for icy particles with compacted-frost surfaces (Hatzes et al., 1988). The outer ring is considered to be confined by the 3:1 mean motion resonance with Quaoar's spin rate and the 6:1 mean motion resonance with Weywot, a satellite of Quaoar (Morgado et al., 2023).

Various formation mechanisms for asteroidal ring systems have been proposed. They broadly fall into four basic concepts and combinations of them. First is those related to impacts (Pan & Wu, 2016; Melita et al., 2017; Ortiz et al., 2017). For Chariklo's ring system, Pan & Wu (2016) and Melita et al. (2017) mentioned that it is physically possible to transfer mass to the orbit through impacts, although the probability of such events to occur is generally low in the present trans-neptunian region. Second is those making use of cometary activities on the surface; Pan & Wu (2016) showed that CO outgassing could eject particles under the thermophysical and chemical condition of Centaurs. Third is those via tidal disruption. Pan & Wu (2016) explored a possibility that close encounter with Neptune can induce the inward migration and subsequent tidal disruption of a satellite. They concluded that it is possible, but the initial configuration of the satellite and the strength of close encounter have to be favorable for this scenario. Melita et al. (2017) showed that when a satellite is comparable in size with the central body and has a slower spin rate or when it is captured at a favorable distance, it would migrate inward through tidal interaction. Hyodo et al. (2016) showed that tidal disruption of a passing body can result in a ring system when it experiences a close encounter with giant planets. Tidal disruption of a satellite (e.g., Canup, 2010; Leinhardt et al., 2012) or a passing object (e.g., Dones, 1991; Hyodo et al., 2017) has been explored as a possible origin of Saturn's rings. Fourth is those involving fast rotation (Braga-Ribas et al., 2014). This is not plausible for Chariklo (Pan & Wu, 2016) based on its current rotation period of ~ 7 h (Fornasier et al., 2014). On the other hand, Haumea is a fast rotator, with a spin period of ~3.9 h (Lellouch et al., 2010). Hence, particles may be ejected from the surface of Haumea and survive in a region between 2:1 outer Lindblad resonance and the classical Roche limit (Sicardy et al., 2019; Sumida et al., 2020; Noviello et al., 2022).

Stability of asteroidal ring systems has been intensively studied making use of the theory of ring dynamics established for planetary rings (e.g., Murray & Dermott, 1999; Esposito, 2014 for reviews). (i) Collisions between particles and subsequent dissipation of energy spread a ring, which is called viscous spreading. Viscous spreading rapidly disperses a ring. In the case of Chariklo's rings, its timescale is estimated to be on the order of ~$10^3$-$10^6$ years although the timescale depends on ring particle sizes and their surface density (Braga-Ribas et al., 2014; Pan & Wu, 2016). Thus, asteroidal ring systems should be actively replenished or very young unless other mechanisms are at work stabilizing. Global *N*-body simulations on Chariklo's ring system show that structures, called self-gravitational wakes, should develop in the inner ring (Michikoshi & Kokubo, 2017). This plays a role in facilitating viscous spreading, and the expected lifetime of the inner ring is on the order of 100 years for meter-sized ring particles (Michikoshi & Kokubo, 2017). In order to stabilize the ring against significantly rapid viscous spreading, shepherding satellites are assumed to be at work (Braga-Ribas et al., 2014; Pan & Wu, 2016; Michikoshi & Kokubo, 2017; Giuliatti Winter et al., 2023; Sickafoose & Lewis, 2024). They are also favorable for confining the observed sharp edges of asteroidal rings (Goldreich & Tremaine, 1982). Note that such satellites have not been detected yet possibly because of their small size (Braga-Ribas et al., 2014). (ii) The evolution of ring systems is also affected by external disturbance. Ring particles, particularly sub-centimeter particles, are susceptible to Poynting-Robertson effect and disperse on the order of less than a few Myr (e.g., Braga-Ribas et al., 2014). (iii) The dynamics of particles is highly affected by the non-axisymmetric gravitational potential from the non-spherical shape, the heterogeneous internal structure, or even relatively minor topographic features of the central body (Sicardy et al., 2019; Winter et al., 2019; Marzari, 2020; Sanchez et al., 2020; Sumida et al., 2020; Madeira et al., 2022; Giuliatti Winter et al., 2023; Ribeiro et al., 2023), and from single or multiple satellites (e.g., Rodríguez et al., 2023; Sickafoose & Lewis, 2024). Sicardy et al. (2019) demonstrated that the non-axisymmetric gravitational potential induces strong resonances and promotes particle migration. (iv) Close encounters with giant planets can disturb asteroidal ring systems. Araujo et al. (2016) and Wood et al. (2017) investigated the orbital evolution of Chariklo by integrating numerous clones with Chariklo-like orbit, and found that Chariklo is unlikely to have experienced or will experience disruptive close encounters with giant planets. Similarly, Araujo et al. (2018) and Wood et al. (2018a) showed that the probability of disruptive close encounters is low for Chiron. Haumea and Quaoar are also implausible to experience close encounters with giant planets, because unlike Centaurs, they reside far beyond Neptune. Hence, these objects are unlikely to undergo extremely close encounters with giant planets that could heavily damage their rings.

In this study, we evaluate the response of asteroidal ring systems when they experience gravitational disturbance by a close encounter with a small object. Collisions serve as an important evolutionary process even in the outer solar system, as presented by the cratered surfaces of Pluto, Charon, and Arrokoth (Singer et al., 2019; Spencer et al., 2020). Since close encounters are more likely to take place than collisions, every small object could experience close encounters with other small objects that may affect the eccentricity of ring particles. We focus on the immediate response to the perturbation, and not on the long-term stability and fate of the perturbed ring. Although previous works addressed the stability of asteroidal ring systems under close encounters with giant planets, we focus on close encounters with small objects, which is expected to take place more frequently. The structure of this paper is as follows. We present the numerical method and parameters in Section 2. In Section 3, we show how each parameter affects the resultant gravitational disturbance of ring systems. In Section 4, we estimate the lifetime of ring systems by computing the number of disruptive close encounters, and explore the dependence on the size of the central body. The definition of symbols used in this study is given in Table 1.

**Table 1.** List of symbols.

| Symbol | Quantity | Value | Unit |
|---|---|---|---|
| $\boldsymbol{r}$ | Position vector of a ring particle with origin at the center of the central body, and $\boldsymbol{r}^* = \boldsymbol{r}/R$. | | m |
| $\boldsymbol{r_p}$ | Position vector of the perturber with origin at the center of the central body, and $\boldsymbol{r_p^*} = \boldsymbol{r_p}/R$. | | m |
| $R$ | Radius of the central body. | Table 2 | m |
| $R_p$ | Radius of the perturber. | | m |
| $R_{p,c}$ | Radius of the perturber equivalent to $M_{p,c}^*$. | | m |
| $R_{\text{ring}}$ | Ring radius. | Table 3 | m |
| $M$ | Mass of the central body. | Table 2 | kg |
| $M_p$ | Mass of the perturber. | | kg |
| $M_p^*$ | Mass ratio of the perturber to the central body ($M_p/M$). | $2^{-6}$-$2^3$ | |
| $M_{p,c}^*$ | Critical mass ratio, with which perturber can cause $e_{\max} = 0.1$ (Section 2.3). | | |
| $M_{p,\max}$ | Maximum mass of the perturber. | 5 Pluto mass | kg |
| $t$ | Time. The scaled time is $t^* = t/T$. | | s |
| $T$ | Scaling parameter of time, $\sqrt{R^3/GM}$. | | s |
| $T_{\text{ring}}$ | Orbital period of ring particles, $T_{\text{ring}} = \sqrt{R_{\text{ring}}^3/GM}$. | | s |
| $T_e$ | Characteristic timescale of the perturbation, $T_e = D/v_{\text{peri}}$. | | s |
| $D$ | Pericenter distance of the perturber's orbit (Fig. 1). In order to avoid collision between the central body and perturber, the minimum pericenter distance is $R + R_p$. | ≤50 | m |
| $D_{max}$ | Maximum pericenter distance, where the critical mass ratio is equal to $M_{p,max}$. | | m |
| $D_{td}$ | Tidal disruption distance from the central body (Eq. 3). | | R |
| $\delta e$ | Change in eccentricity of ring particles after close encounter. | | |
| $e_{max}$ | Maximum $\delta e$. | | |
| $f$ | Fraction of lost ring particles to a total after close encounter. | 0-1 | |
| $i_p$ | Inclination of the perturber's orbit. | 0-180 | degree |
| $\omega_p$ | Argument of pericenter of the perturber's orbit. | 0-90 | degree |
| $v_{peri}$ | Velocity at pericenter of the perturber relative to the central body. | 2.0 | km s$^{-1}$ |
| $\alpha, \beta, \gamma, \kappa$ | Fitting parameters (Section 3.1). | | |
| $N(<D)$ | The number of close encounters that occur within $D$ and cause $e_{max} > 0.1$. | | yr$^{-1}$ |
| $\tau(<D)$ | Lifetime of the ring system, $\tau(<D) = 1/N(<D)$. | | yr |
| $\tau_{ring}$ | Lifetime of the ring system for $D = D_{max}$. | | yr |

## 2. Method

### 2.1. Numerical simulation

We suppose a close encounter between a ringed object (hereafter referred to as the central body) and a perturber (i.e., another object passing by the central body). Solar perturbations are not included in our simulations. Figure 1 shows the encounter geometry. We take the origin of the coordinate system $(X, Y, Z)$ at the center of the central body. We assume a narrow and circular ring with a radius of $R_{ring}$ on the $X$-$Y$ plane. The ring consists of 10 annuli and we place 500 particles evenly on each annulus, which makes 5000 particles in total. The initial eccentricity and inclination of the ring particles are 0. We define $r$ and $r_p$ as the position vectors of a ring particle and the perturber with respect to the central body, respectively, $M$ and $M_p$ as the masses of the central body and perturber, and $R$ and $R_p$ as the radii of the central body and perturber. We consider the gravitational effect of the perturber on each particle in the ring. In order to solve the motion of a ring particle, we integrate the following equation of motion (Brouwer & Clemence, 1961; Murray & Dermott, 1999):

$$\frac{d^2 r}{dt^2} = -\frac{GM}{|r|^3} r - \frac{GM_p}{|r-r_p|^3}(r-r_p) - \frac{GM_p}{|r_p|^3} r_p, \quad (1)$$

where $G$ is the gravitational constant. The first term on the r.h.s. is the gravitational force between the particle and the central body, and the second and third terms represent the direct and indirect perturbing forces, respectively, from the passing object (perturber). Scaling mass by the mass of the central body, $M$, length by the radius of the central body, $R$, and time by $T = \sqrt{R^3/GM}$, we obtain the scaled equation of motion:

$$\frac{d^2\boldsymbol{r}^*}{dt^{*2}} = -\frac{\boldsymbol{r}^*}{|\boldsymbol{r}^*|^3} - M_p^*\frac{\boldsymbol{r}^* - \boldsymbol{r}_p^*}{|\boldsymbol{r}^* - \boldsymbol{r}_p^*|^3} - M_p^*\frac{\boldsymbol{r}_p^*}{|\boldsymbol{r}_p^*|^3}, \quad (2)$$

where $\boldsymbol{r}^*$ is the scaled position vector of a ring particle ($\boldsymbol{r}^* = \boldsymbol{r}/R$), $\boldsymbol{r}_p^*$ is the scaled position vector of the perturber ($\boldsymbol{r}_p^* = \boldsymbol{r}_p/R$), $M_p^*$ is the scaled mass of the perturber ($M_p^* = M_p/M$), and $t^*$ is the scaled time ($t^* = t/T$). The trajectory of the perturber is hyperbolic about the central body. The initial position of the perturber is set at $|\boldsymbol{r}_p^*| = 100$. We define the origin of time so that the closest approach takes place at $t = t^* = 0$. Calculations are terminated at $t^* = 10T_{\text{ring}}^*$, where $T_{\text{ring}}^* = T_{\text{ring}}/T$ and $T_{\text{ring}}$ is orbital period of ring particles ($T_{\text{ring}} = \sqrt{R_{\text{ring}}^3/GM}$). We confirmed that, at the end of each calculation, the perturber is sufficiently far away from the ring and thus the total energy of the ring particles is no longer affected by the perturber. We run our simulation using the 4th-order Runge-Kutta integrator. We set a time step of $\Delta t^* = 2^{-10}$, which corresponds to less than a few seconds in unscaled time. Face-on snapshots in Fig. 2 illustrate an example of the temporal evolution in the case of Haumea's ring.

At the end of each calculation, we obtain the change in eccentricity of each ring particle, $\delta e$, from the position and velocity of each ring particle. We define $e_{\max}$ as the maximum value of $\delta e$ of all ring particles in each calculation. We also obtain the fraction of lost particles (i.e., the ratio of ring particles lost to a total), $f$. We define that a ring particle is lost when (i) during the calculation, it collides with the central body ($|\boldsymbol{r}^*| \leq 1$) or the perturber ($|\boldsymbol{r}^* - \boldsymbol{r}_p^*| \leq R_p/R$), or reaches beyond the Hill radius of the central body with regard to the Sun (on the order of $10^3 R$ for Chariklo, Haumea, and Quaoar), (ii) at the end of the calculation, the pericenter or apocenter of its osculating orbit is within the radius of the central body or beyond the Hill radius, or (iii) at the end of the calculation, the eccentricity of its orbit is larger than 1 ($\delta e \geq 1$). We set the density of the perturber to 1000 kg/m³ to define $R_p$ (e.g., Parker & Kavelaars, 2012; Nesvorný et al., 2021; Campbell et al., 2023). Note that the perturber's radius is almost the same (only 0.8 times smaller) even if we use 2000 kg/m³. We do not consider the tidal deformation of both the central body and perturber and thus their radii remain constant.

In this study, we utilize the change in eccentricity as the measure of the severity of close encounters, rather than that in inclination. This is because $\delta e$ is a key factor for the fate of ring particles: a significant change in eccentricity may result in particle loss from the system owing to collision or ejection. In contrast, the change in inclination does not affect the fate of disturbed particles in short-term evolution, as we do not take into account interparticles collisions and non-axisymmetric gravitational potential. The long-term evolution of eccentricity and inclination after disturbance is beyond the scope of this study. Moreover, it has become customary in the literature to characterize the damage of a ring in terms of the eccentricity of ring particles (e.g., Araujo et al., 2016, 2018; Wood et al., 2017, 2018a, 2018b). In any case, we show that the eccentricity of ring particles is hardly changed by close encounters (Sections 3, 4), and hence our conclusion should be unaffected by the change in inclination of ring particle.

## 2.2. Parameter space

We study cases of the inner ring of Chariklo, the ring of Haumea, and the outer ring of Quaoar listed in Table 2 and 3. We ignore the outer ring of Chariklo because (i) the gap between the inner and outer ring is very small and (ii) the inner ring is wider and denser. We used 7 km as the width of Chariklo's inner ring (Braga-Ribas et al., 2014; Morgado et al., 2021). We also ignore the inner ring of Quaoar because the outer ring is denser and more prone to the perturbation than the inner ring. Although Quaoar's outer ring is inhomogeneous (Morgado et al., 2023), we here regard it as a ring with a constant width of 5 km where the highest optical depth was observed.

The velocity of the perturber at pericenter, $v_{\text{peri}}$, is fixed at 2.0 km/s based on the impact velocity that hot classical TNOs are most likely to experience (Abedin et al., 2021). Although the classification of TNOs varies between groups of authors, we assume that Haumea and Quaoar fall into hot classical TNOs since their orbital inclinations are greater than 5° (Table 2) (Gladman et al., 2008; Abedin et al., 2021). In this study, we use the same parameters for Chariklo, although it is not a hot classical TNO. Note that Centaurs are considered to be originally dynamically excited TNOs (e.g., Levison & Duncan, 1997).

## 2.3. Critical mass ratio

In order to estimate the lifetime of a ring system from the perspective of a close encounter with another small object, we need to define which close encounters are critical. Previous studies evaluated the severity of a close encounter upon a binary system by various means including pericenter distance, change in eccentricity ($e_{\text{max}}$), change in energy, or combinations of them (e.g., Chauvineau et al., 1991; Araujo & Winter, 2014; Araujo et al., 2016, 2018; Wood et al., 2017, 2018a, 2018b). Pericenter distance is a useful criterion to investigate the severity of close encounters with constant masses like giant planets. As shown in Section 3, perturber mass, as well as pericenter distance, significantly affects the severity in our study. Araujo et al. (2016) assumed $e_{\text{max}} \gg 0.01$ to be a noticeable disturbance. Following this previous study, we define the critical mass ratio, $M_{p,c}^*$, as the mass ratio of a perturber that can cause $e_{\text{max}} = 0.1$. In other words, we assume that a ring is significantly disturbed when a close encounter takes places with a perturber whose mass ratio is greater than this critical mass ratio. We confirmed that this criterion is consistent with the relationship between $e_{\text{max}}$ and $f$

(Section 3.3). Because $M_{p,c}^*$ depends on pericenter distance, $D$, we obtain $M_{p,c}^*$ at each $D$ by using the relationships between $e_{\max}$ and $M_p^*$ (Section 3.3).

## 2.4. Analytical expressions

Investigations on stellar encounters with protoplanetary disks are carried out to evaluate the effect on planets (Lyttleton & Yabushita, 1965), planetesimal disk (Kobayashi & Ida, 2001; Larwood & Kalas, 2001), Kuiper belt objects (KBOs) (e.g., Ida et al., 2000; Levison et al., 2004; Kobayashi et al., 2005), and extrasolar planets (Zakamska & Tremaine, 2004). Furthermore, close encounters are considered to play an essential role in the evolution of stellar binaries (e.g., Yabushita, 1966; Heggie & Rasio, 1996) and binary asteroids (e.g., Chauvineau et al., 1991; Walsh & Richardson, 2006). Those works theoretically obtained analytical expressions of the change in eccentricity, $\delta e$, under close encounters; for example, (i) Heggie & Rasio (1996) obtained analytical expressions using the Laplace-Runge-Lenz vector, (ii) Kobayashi & Ida (2001) an analytical expression by using Gauss's method (Brouwer & Clemence, 1961; Murray & Dermott, 1999), and (iii) Zakamska & Tremaine (2004) analytical expressions from the impulse approximation and the Laplace-Lagrange secular theory (Brouwer & Clemence, 1961; Murray & Dermott, 1999).

When the orbital period of ring particles, $T_{\mathrm{ring}}$, is much shorter than the characteristic timescale of the perturbation, $T_e$ (Zakamska & Tremaine, 2004), ring particles can complete their orbits during encounter. Then, ring particles receive disturbance from the perturber at the entire orbit, which leads their perturbed orbital elements to be averaged over the whole orbit (orbital averaging). Orbital averaging reduces the short-term variation in orbital elements and highlights the effect of secular or long-term variation. The Laplace-Lagrange secular theory is applicable, when orbital averaging is valid and the change in eccentricity and inclination of ring particles after encounter is sufficiently small (Brouwer & Clemence, 1961; Murray & Dermott, 1999). Zakamska & Tremaine (2004) obtained expressions with additional assumptions that the trajectory of a perturber is nearly straight (ZT04Secular), or not straight (ZT04Slow). Kobayashi & Ida (2001) developed an analytical expression (KI01) using another secular theory called Gauss's method. KI01 assumed that the change in eccentricity and inclination of ring particles after encounter is sufficiently small and orbital averaging is valid. Zakamska & Tremaine (2004) additionally obtained an analytical expression using the impulse approximation (ZT04Imp), which assumes $T_e/T_{\mathrm{ring}} \ll 1$ (i.e., the relative motion of ring particles during an encounter is negligible). On the other hand, using Laplace-Runge-Lenz vector, Heggie & Rasio (1996) obtained analytical expressions (HR96Exp) when orbital averaging is not valid, which is suitable for encounters that take place at relatively small pericenter distance and/or with slow perturbers. They also obtained an analytical expression when orbital averaging is valid (HR96Power). Both

Kobayashi & Ida (2001) and Zakamska & Tremaine (2004) noted that their expressions are equivalent with HR96Power.

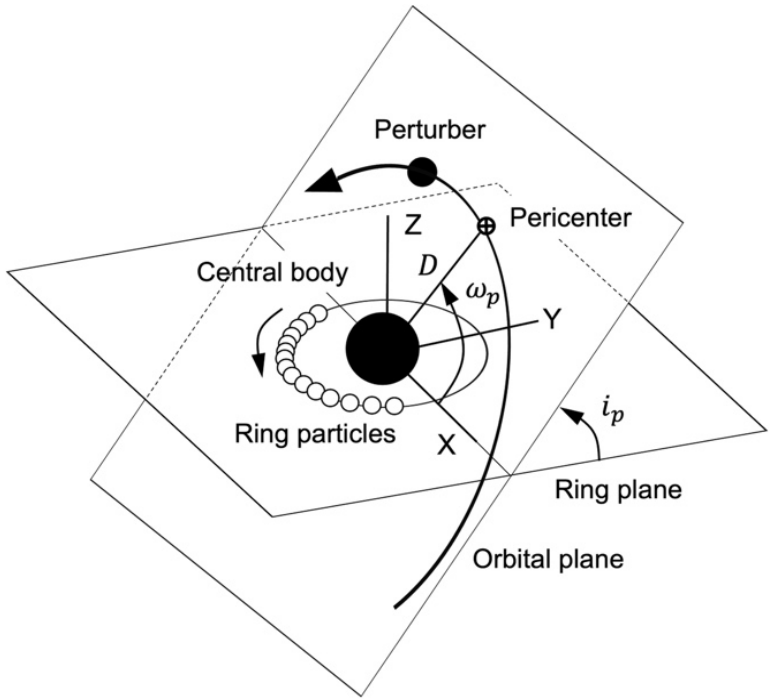

**Fig. 1.** Encounter geometry: the central body (large filled circle), perturber (small filled circle), ring particles (small open circles), and the pericenter of perturber orbit (crossed circle).

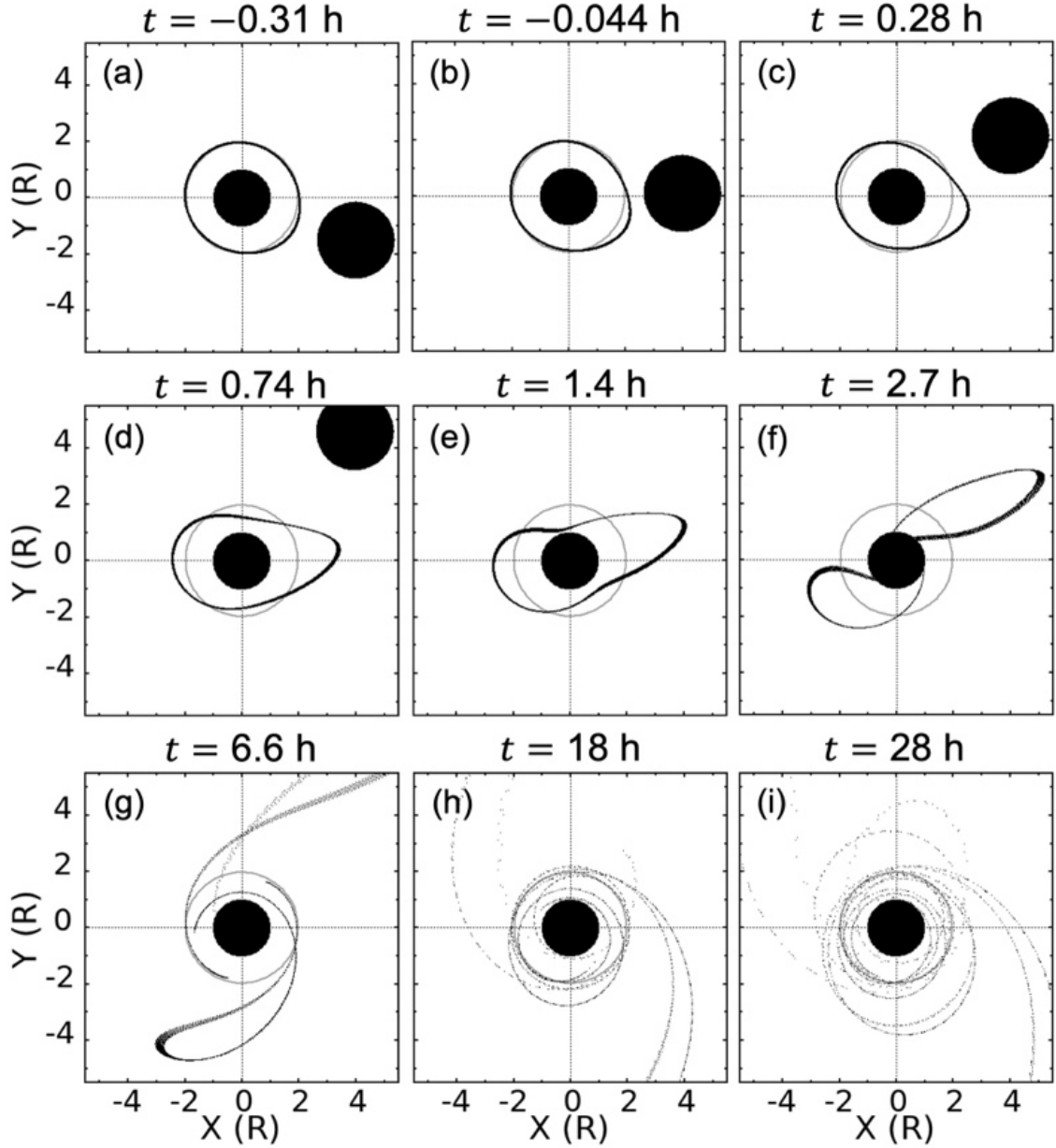

**Fig. 2.** Face-on snapshots of temporal evolution, from (a) to (i), of Haumea's ring at a close encounter assuming $M_p^* = 4$, $D = 4R$, $i_p = 0°$, and $\omega_p = 0°$. The black filled circle at the origin is Haumea (the central body) and the other is the perturber. The black circle in panel (a) is Haumea's ring and consists of dots that represent each ring particle. The gray circle with a radius of $2R$ is the initial location of the ring. Closest approach takes place at $t = 0$ and the time is shown in the unit of hours.

**Table 2.** The orbital and physical properties of Chariklo, Haumea, and Quaoar.

| | Radius | Mass | Eccentricity | Inclination | Semi-major |
|---|---|---|---|---|---|

|  | (km) | (kg) | c | (deg) c | axis (au) c |
|---|---|---|---|---|---|
| **Chariklo** | 124 ± 9 [a] | 7.986 × 10$^{18}$ [b] | 0.168 | 23.352 | 15.8 |
| **Haumea** | 1161 ± 30 [d] | 4.006 × 10$^{21}$ [e] | 0.200 | 28.211 | 42.9 |
| **Quaoar** | 555 ± 2.5 [f] | 1.4 × 10$^{21}$ [g] | 0.04 | 7.99 | 43.5 |

[a] Braga-Ribas et al. (2014)

[b] Araujo et al. (2016)

[c] JPL small-body database, https://ssd.jpl.nasa.gov/tools/sbdb_lookup.html

[d] Ortiz et al. (2017)

[e] Ragozzine & Brown (2009)

[f] Braga-Ribas et al. (2013)

[g] Fraser et al. (2013)

**Table 3.** The properties of rings of Chariklo, Haumea, and Quaoar.

|  | **Ring radius (km)** | **Ring radius** | **Ring width (km)** |
|---|---|---|---|
| **Chariklo's inner ring** | 390.6 ± 3.3 [a] | 3.1 $R_{\text{Chariklo}}$ [a] | 6.16-7.17 [a]<br>4.8-9.1 [b] |
| **Chariklo's outer ring** [a] | 404.8 ± 3.3 | 3.2 $R_{\text{Chariklo}}$ | 3.4-3.6 |
| **Haumea's ring** [c] | 2287 | 2.0 $R_{\text{Haumea}}$ | 70 |
| **Quaoar's inner ring** [d] | 2520 | 4.6 $R_{\text{Quaoar}}$ | 10 |
| **Quaoar's outer ring** [e] | 4100 | 7.4 $R_{\text{Quaoar}}$ | 4.91-336.34 |

[a] Braga-Ribas et al. (2014)

[b] Morgado et al. (2021) reported a mean value of 6.9 km.

[c] Ortiz et al. (2017)

[d] Pereira et al. (2023)

[e] Morgado et al. (2023)

## 3. Results

### 3.1. Dependence on mass ratio and pericenter distance

In this subsection, we explore the dependence of $\delta e$ and $e_{\max}$ on $M_p^*$ and $D$,

assuming $i_p = 0°$ and $\omega_p = 0°$ as constants. Figure 3 shows the distributions of $\delta e$ as a function of $M_p^*$ in the vicinity of the ring. As a result, Chariklo's ring requires a perturber with larger $M_p^*$ than that of the other two rings for the similar degree of disturbance; for example, a perturber of $M_p^* = 2^{-6}$ can raise $e_{max}$ to $10^{-3}$ for Chariklo's ring, while $e_{max}$ to $10^{-2}$ for Haumea's and Quaoar's ring. Although Quaoar's ring lies much farther than Haumea's (in unit of their radius), Quaoar's ring requires a perturber with similar $M_p^*$ to bring about the similar degree of disturbance with Haumea's ring. Resultant $\delta e$ and $e_{max}$ for all three rings follow a power-law of $M_p^*$. We fit the value of $e_{max}$ to a power-law of the form, $e_{max} \propto (M_p^*)^\alpha$, and the best-fit parameter, $\alpha$, as a function of $D$ is shown in Fig. 4. We find that $\alpha \sim 1$ for all three bodies when $D > R_{ring}$ and $\alpha \sim 0.7$ when $D \leq R_{ring}$. This is consistent with all analytical expressions mentioned in Section 2.4 that predict $\delta e$ and $e_{max}$ following a linear function ($\delta e, e_{max} \propto M_p^*$).

Figure 5 shows the distribution of $\delta e$ as a function of $D$. The dispersion of $\delta e$ is approximately a few orders of magnitude when $D$ is small (closer to the ring) and becomes smaller as $D$ increases (farther from the ring). Chariklo's and Quaoar's rings show a dispersion of $\delta e$ about four orders of magnitude at $D = 10R$ and one order of magnitude at $D = 50R$ (Fig. 5a, d). The dispersion of $\delta e$ for Haumea's ring rapidly narrows down and is much smaller than one order of magnitude at $D = 50R$ (Fig. 5b, c). The dispersion of $\delta e$ is not dependent on $M_p^*$ (Fig. 3) but on $D$. This is because as the perturber passes close to the ring (i.e., $D$ is small), ring particles experience a different degree of disturbance depending on the relative position with the perturber. For Chariklo's ring, a sharp depression is seen at $D = 10R$ (Fig. 5a). For Quaoar's ring, a similar depression is also observed at $D = 10R$ when $M_p^*$ is small (Fig. 5d), while it gradually becomes less prominent as $M_p^*$ increases (Fig. 5e) and disappears for $M_p^* \geq 2$ (Fig. 5f). No such depression is observed for Haumea's ring. We find that $e_{max}$ for Chariklo's or Quaoar's ring can be approximated by a power-law of $D$. For Haumea's ring, $e_{max}$ does not match a power-law but follows an exponential function. We fit the value of $e_{max}$ to a power law of the form, $e_{max} \propto D^\beta$, for Chariklo's and Quaoar's rings, and an exponential function of the form, $e_{max} \propto \exp(-\kappa D^\gamma)$, for Haumea's ring. The best-fit parameters, $\beta$, $\kappa$, and $\gamma$, as a function of $M_p^*$ are shown in Fig. 6. The power law index, $\beta$, varies from -2.2 to -2 for Chariklo's ring, and from -2.4 to -2.8 for Quaoar's ring. For Haumea's ring, $e_{max}$ approximates an exponential function, $e_{max} \propto \exp(-0.90 D^{0.56})$ when $M_p^* \leq 1$. As $M_p^*$ increases to 8, $\kappa$ increases to 1.6 and $\gamma$ decreases to 0.46.

We find that the change in eccentricity of Chariklo's and Quaoar's rings in our calculations agrees with ZT04Imp, while that of Haumea's ring with HR96Exp or ZT04Imp. For Chariklo's and Quaoar's rings, $T_{ring}$ is roughly 20 and 50 hours, respectively, leading to $T_e/T_{ring}$ on the order of $10^{-2}$ when $D = 50R$. Therefore, it is reasonable that ZT04imp agrees with the case of Chariklo's and Quaoar's rings (Figs. 3 and 5). In the case of Haumea's ring, the behavior of $\delta e$ follows HR96Exp especially when $M_p^*$ is small and $D$ is large (Fig. 5b, c). This is presumably because $T_e$ is comparable to $T_{\text{ring}}= 11$ hours ($T_e/T_{\text{ring}} \sim 0.7$ at $D = 50R$) and is not large enough for the impulse approximation to be valid. As $M_p^*$ increases and $D$ decreases, the deviation from HR96Exp becomes large and $\delta e$ follows ZT04imp (Figs. 3b and 5b, c). We consider that this transition takes place because $T_e/T_{\text{ring}} \ll 1$ is gradually achieved for smaller $D$ ($T_e/T_{\text{ring}} \sim 0.1$ at $D = 10R$).

Figure 7 shows the fraction of lost particles, $f$, as a function of $M_p^*$ and $D$, assuming $i_p = 0°$, and $\omega_p = 0°$. For Chariklo's ring, in the case of $M_p^* \leq 8$, particle loss in the criteria in Section 2.1 does not take place ($f = 0$) if $D > R_{\text{ring}} + R_p$. For Haumea's ring, (i) in the case of $M_p^* \leq 1/4$, particle loss does not take place unless $D \leq R_{\text{ring}} + R_p$, (ii) in the case of $1/4 \leq M_p^* < 1$, all of the lost ring particles are eliminated by either collision with the central body or perturber, and (iii) in the case of $M_p^* \geq 1$, some fraction of the lost ring particles is eliminated by not only collision but also escape from the system ($\delta e \geq 1$). As $M_p^*$ increases and $D$ decreases, larger fraction of the lost ring particles is eliminated by escape from the system. For example, when $M_p^* = 8$ and $D = 3R$ for Haumea's ring, the fractions of ring particles eliminated by collision with the perturber, by collision with the central body, and by escape from the system are approximately equal. On the other hand, when $M_p^* = 2$ and $D = 3R$, 80% of the lost ring particles are eliminated by collision with the central body or perturber and 20% is eliminated by escape from the system. A few particles, typically one or two, are eliminated because their apocenter is beyond the Hill radius at the end of the calculation. For Quaoar's ring, (i) in the case of $M_p^* < 1/4$, particle loss does not take place if $D > R_{\text{ring}} + R_p$ and (ii) in the case of $M_p^* \geq 1/4$, some fraction of the lost ring particle is eliminated either by collision with the central body or by escape from the system, although the majority of the lost ring particles is eliminated by collision with the perturber. Escape from the system predominates over collision with the central body in the case of Quaoar's ring than that of Haumea's ring, because Quaoar's ring lies farther from the central body than Haumea's. Only one or two particles, but several particles for large $M_p^*$, are eliminated because the pericenter or apocenter of their orbit is within the central body or beyond the Hill radius at the end of the calculation.

Agnor & Hamilton (2006) considered the Hill radius of a binary system and provided an equation for tidal disruption distance, $D_{\text{td}}$, which can be used as a theoretical threshold for binary breakup. Since ring systems in this study can be regarded as a binary consisted of the central body and a massless ring particle, we obtain $D_{\text{td}}$ by

$$D_{\text{td}} = R_{\text{ring}}(3M_p^*)^{\frac{1}{3}}. \tag{3}$$

Figure 7 shows that Eq. (3) can provide a rough estimate for the pericenter distance where ring particles start to be lost despite deviation of a few $R$.

### 3.2. Dependence on orbital elements

Left panels of Fig. 8 show the distribution of $e_{\max}$ as a function of $i_p$ and $\omega_p$ in the vicinity of the ring assuming $M_p^* = 1$. For all three rings, we found that $e_{\max}$ varies approximately one order of magnitude depending on the combination of $(i_p, \omega_p)$, and $e_{\max}$ is slightly larger near $i_p = 0°$ (prograde) than near $i_p = 180°$ (retrograde). For Chariklo's and Quaoar's rings, the maximum $e_{\max}$ is near $(i_p, \omega_p) = (90°, 0°)$. For Haumea's ring, the maximum $e_{\max}$ is near $i_p = 0°$. The value of $e_{\max}$ gradually decreases toward $(i_p, \omega_p) = (90°, 90°)$ in the case of Quaoar's ring. On the other hand, there are a few local minima near $(i_p, \omega_p) = (30°, 90°)$, $(90°, 90°)$, and $(150°, 90°)$ in the case of Chariklo's and Haumea's rings. Although $e_{\max}$ takes similarly small values at both $(i_p, \omega_p) = (30°, 90°)$ and $(150°, 90°)$ in the case of Chariklo's ring, $e_{\max}$ takes much smaller value at $(i_p, \omega_p) = (150°, 90°)$ than the other two minima in the case of Haumea's ring. Both the impulse approximation and the secular approximation predict that $\delta e$ is independent of the direction of the perturber's motion relative to the rings' orbital motion (Kobayashi & Ida, 2001; Zakamska & Tremaine, 2004), which is responsible for the nearly symmetric profiles about $i_p = 90°$ for Chariklo's and Quaoar's rings (Fig. 8a, c). On the other hand, such a symmetric profile is not seen for Haumea's ring (Fig. 8b), which indicates that the close encounter is not in the regime of both the impulse approximation and the secular approximation. Right panels of Fig. 8 illustrate the dependence of $f$ on $i_p$ and $\omega_p$ in the vicinity of the ring. For Chariklo's and Quaoar's rings, particles are lost when $i_p$ is close to $0°$ or $180°$ or when $\omega_p$ is close to $0°$. This is because collision with the perturber is the primary cause of particle loss. For Haumea's ring, particles are lost when $i_p \leq 30°$ with wide range of $\omega_p$ or when $i_p \leq 120°$ and $\omega_p \leq 30°$. In the case of Haumea's ring, particles are lost through collision with the central body as well as that with the perturber.

Hence, the largest $f$ and $e_{\max}$ can be expected when $i_p = 0°$ and $\omega_p = 0°$. This means that $M_{p,c}^*$ with $(i_p, \omega_p) = (0°, 0°)$ can be regarded as the minimum $M_{p,c}^*$ at given $D$ to cause $e_{\max} = 0.1$ (Section 3.3). Since the smaller objects are more abundant, we can estimate the minimum lifetime between close encounters to occur (Section 4).

### 3.3. Critical mass ratio

Figure 9 shows the relationship between $e_{\max}$ and $f$ for cases with various set of $(M_p^*, D)$ explored in Section 3.1. Particles start to be lost ($f > 0$) at $e_{\max} \sim 0.5$ for Haumea's ring, and at $e_{max} \sim 0.1$ for Quaoar's ring. Therefore, as mentioned in Section 2.3, we define the critical mass ratio, $M^*_{p,c}$, as mass ratio with which a perturber can cause $e_{max} = 0.1$, although Chariklo's ring particles are only lost when $D \leq R_{ring} + R_p$.

Figure 10 exhibits $M^*_{p,c}$ as a function of $D$. Hereafter, we set $i_p = 0°$, and $\omega_p = 0°$ to obtain the minimum $M^*_{p,c}$ at given $D$ (Section 3.2). Even in the immediate vicinity of the ring, $M^*_{p,c}$ is on the order of $10^{-1}$ for Chariklo's ring and on the order of $10^{-2}$ for Haumea's and Quaoar's rings. Approximately at $10R$ away from the ring, $M^*_{p,c} \geq 1$ is required for Haumea's and Quaoar's rings. For Chariklo's ring, $M^*_{p,c}$ as a function of $D$ approximates,

$$M^*_{p,c} = 0.185 \times D^{1.94}, \quad (4)$$

for Quaoar's ring,

$$M^*_{p,c} = 0.00326 \times D^{2.11}, \quad (5)$$

for Haumea's ring,

$$\begin{cases} M^*_{p,c} = \exp[-9.39D^{-0.952} + 1.39] \text{ for } D \leq 7R, \\ M^*_{p,c} = \exp[0.900D^{0.562} - 2.31] \text{ for } D \geq 7R. \end{cases} \quad (6)$$

We also obtain mass ratio that can yield $e_{max} = 1$ with similar procedure. To cause $e_{max} \geq 1$, a perturber with mass ratio greater than 1 is required even if close encounter takes place very close to the ring. For Chariklo's ring $M^*_{p,c}$ as a function of $D$ approximates,

$$M^*_{p,c} = 1.85 \times D^{1.94}, \quad (7)$$

for Quaoar's ring,

$$M^*_{p,c} = 0.0322 \times D^{2.11}, \quad (8)$$

for Haumea's ring,

$$\begin{cases} M^*_{p,c} = \exp[-9.66D^{-0.999} + 3.59] \text{ for } D \leq 7R, \\ M^*_{p,c} = \exp[0.905D^{0.561} - 0.0169] \text{ for } D \geq 7R. \end{cases} \quad (9)$$

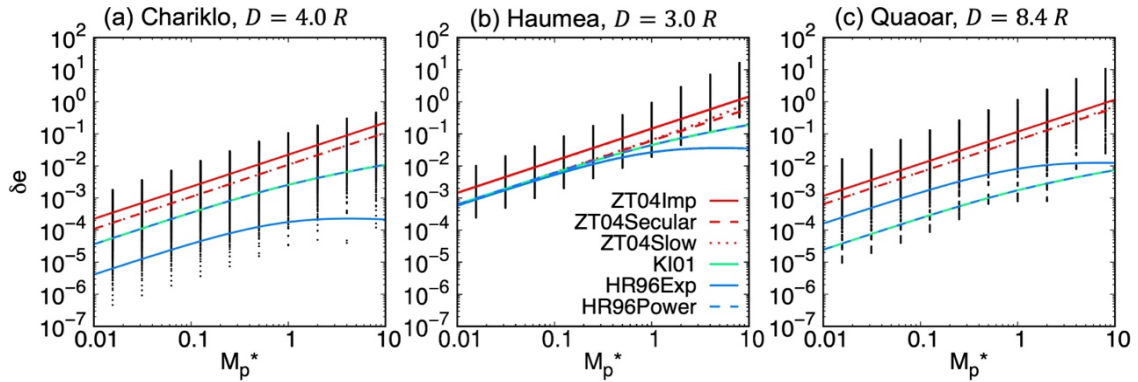

**Fig. 3.** Change in eccentricity, $\delta e$, as a function of $M^*_p$, assuming $i_p = 0°$ and $\omega_p = 0°$. We show results at $D = R_{ring} + R$ (i.e., the perturber passes one central-body radius away from the ring at pericenter). Each black vertical line illustrates the result of each calculation and consists

of small dots representing the pumped-up eccentricity of each ring particle. The top of each black vertical line corresponds to $e_{max}$. The analytical expressions given by Heggie & Rasio (1996) denoted as HR96 (blue lines), Kobayashi & Ida (2001) as KI01 (green line), and Zakamska & Tremaine (2004) as ZT04 (red lines) are shown (see also Section 2.4). Note that HR96Power (dashed blue line) is overlapped by KI01, and ZT04Secular (dashed red line) by ZT04Slow (dotted red line).

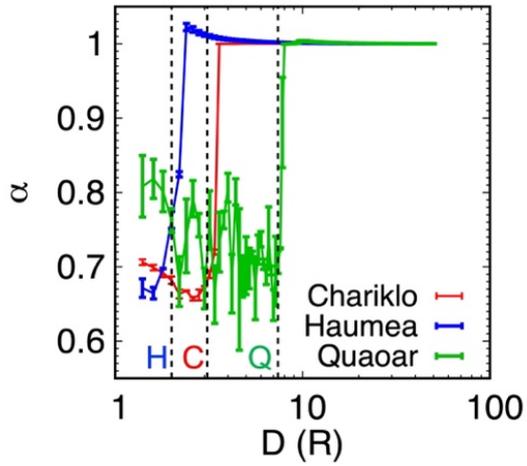

**Fig. 4.** Power law index, $\alpha$, obtained by the power law of the form, $e_{max} \propto (M_p^*)^\alpha$ at given $D$. The vertical black lines indicate the location of the ring. Error bars denote the $\pm 1$ standard deviation.

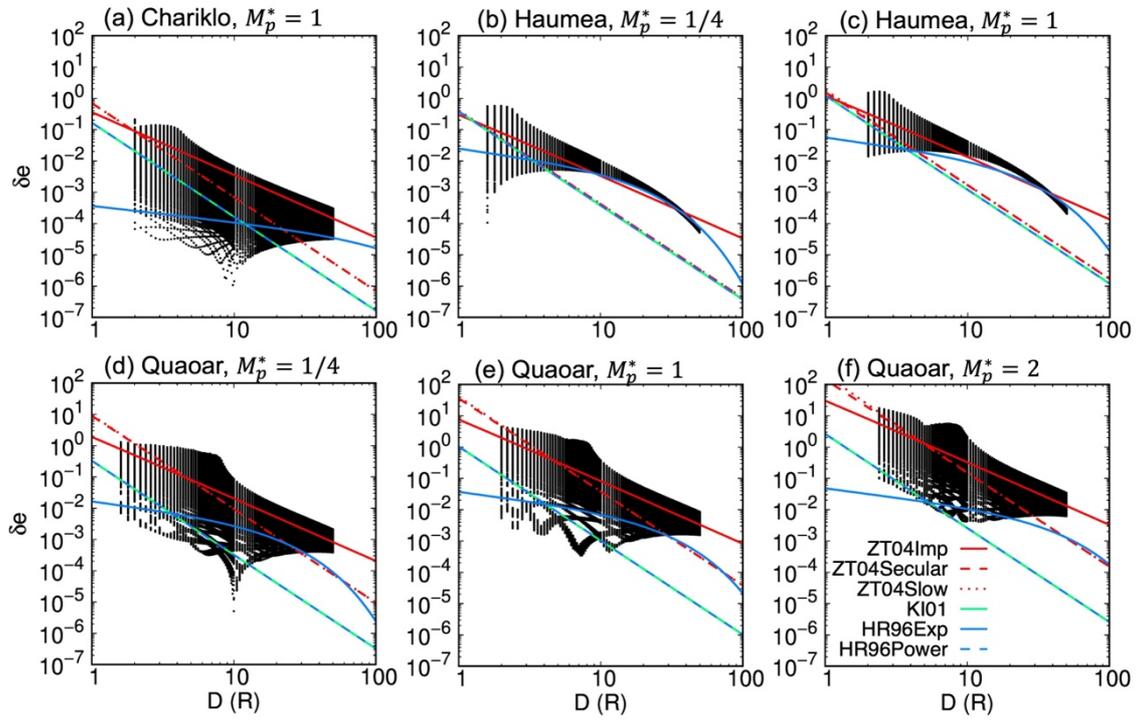

**Fig. 5.** Change in eccentricity, $\delta e$, as a function of $D$ assuming $i_p = 0°$ and $\omega_p = 0°$. (a) Chariklo's inner ring with $M_p^* = 1$, (b-c) Haumea's ring with $M_p^* = 1/4$ and 1, and (d-f) Quaoar's outer ring with $M_p^* = 1/4$, 1, and 2. Each black vertical line illustrates the outcome of each calculation and consists of small dots representing the pumped-up eccentricity of each ring particle. The top of each black vertical line corresponds to $e_{\max}$. The analytical expressions given by Heggie & Rasio (1996) denoted as HR96 (blue lines), Kobayashi & Ida (2001) as KI01 (green line), and Zakamska & Tremaine (2004) as ZT04 (red lines) are shown. Note that HR96Power (dashed blue line) is overlapped by KI01, and ZT04Secular (dashed red line) by ZT04Slow (dotted red line).

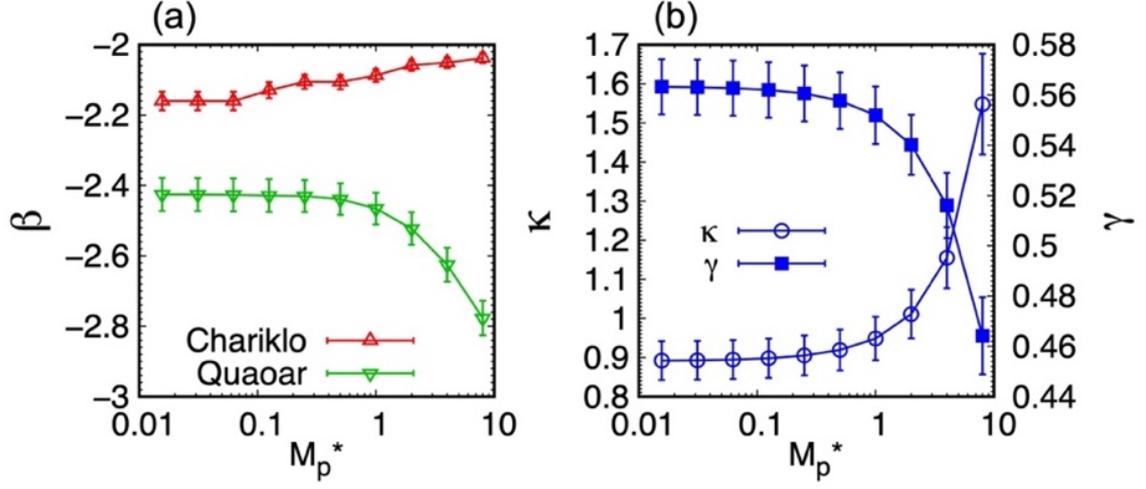

**Fig. 6.** (a) Power law index, $\beta$, obtained through fitting $e_{\max}$ by a power law of the form, $e_{\max} \propto D^{\beta}$ for a given $M_p^*$. (b) Parameters, $\kappa$ and $\gamma$, of Haumea's ring obtained through fitting $e_{\max}$ by an exponential function of the form, $e_{\max} \propto \exp(-\kappa D^{\gamma})$ for a given $M_p^*$. Error bars denote the $\pm 1$ standard deviation.

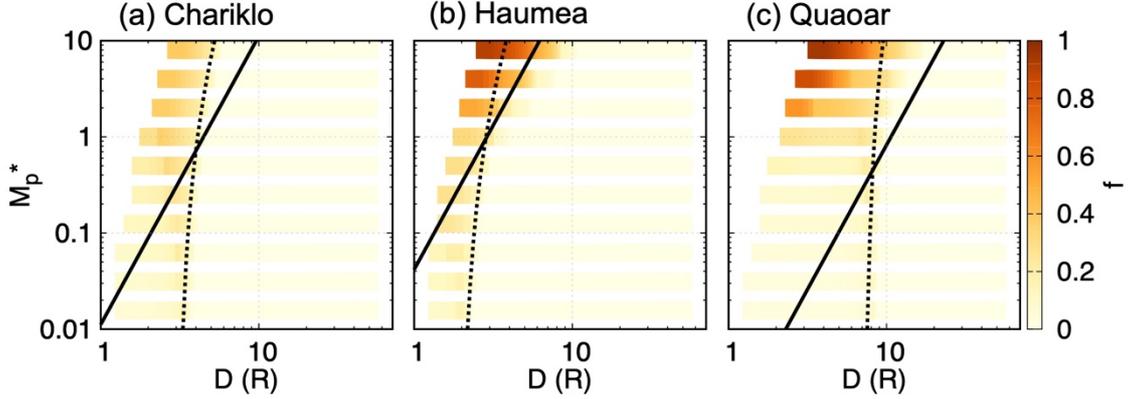

**Fig. 7.** Fraction of lost particles, $f$, as a function of $M_p^*$ and $D$, assuming $i_p = 0°$ and $\omega_p = 0°$ for (a) Chariklo's inner ring, (b) Haumea's ring, and (c) Quaoar's outer ring. The black solid lines indicate $D_{\text{td}}$ given in Eq. (3). The pericenter distance smaller than the black dashed lines shows where $D = R_{\text{ring}} + R_p$. The lack of the results at top-left corner (large $M_p^*$ and small $D$) is because we excluded the cases where the collision between the central body and the perturber occurs ($D \leq R + R_p$).

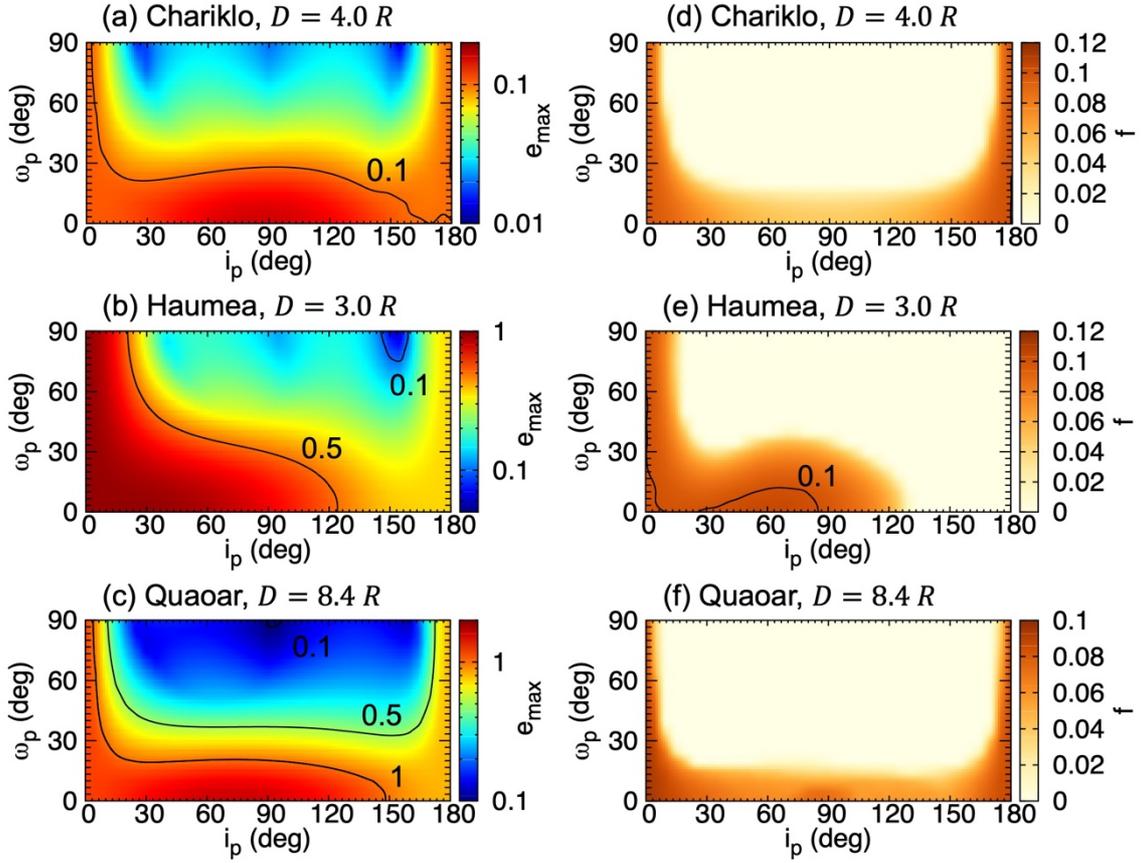

**Fig. 8.** (a-c) The maximum eccentricity, $e_{\max}$, as a function of $i_p$ and $\omega_p$, assuming $M_p^* = 1$. We show results at $D = R_{\text{ring}} + R$ (i.e., the perturber passes one central-body radius away from the ring at pericenter). (d-f) Same with (a-c) but for the fraction of lost particles, $f$.

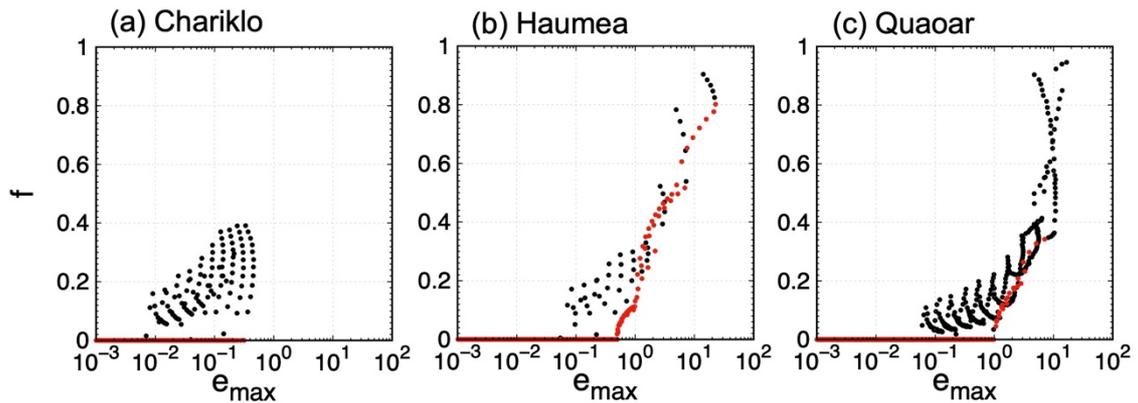

**Fig. 9.** Relationships between the maximum eccentricity, $e_{\max}$, and the fraction of lost particles, $f$ for (a) Chariklo's inner ring, (b) Haumea's ring, and (c) Quaoar's outer ring, assuming $i_p = 0°$ and $\omega_p = 0°$. Red dots represent the cases for $D > R_{\text{ring}} + R_p$, and black dots for $D \leq R_{\text{ring}} + R_p$.

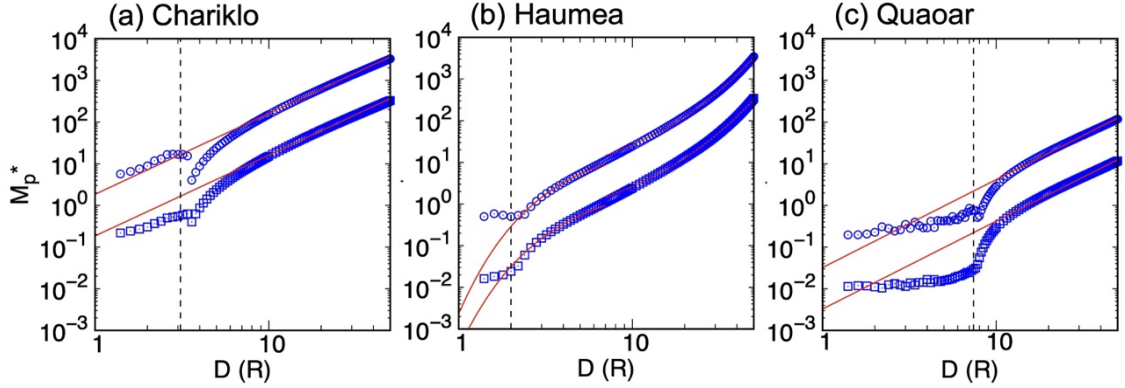

**Fig. 10.** Critical mass ratio, $M_{p,c}^*$, that can cause $e_{\max} = 0.1$ (squares) as a function of $D$ is shown for (a) Chariklo's ring, (b) Haumea's ring, and (c) Quaoar's ring. Mass ratio that can cause $e_{\max} = 1$ is also shown (circles). We assumed $i_p = 0°$ and $\omega_p = 0°$. The vertical black dashed lines indicate the ring radius. Red curves show fitted functions of Eqs. (4)-(9).

## 4. Discussion

### 4.1. The number of close encounters and lifetime

We estimate the lifetime of ring systems in terms of a close encounter with another small object. We define $N(<D)$ as the number of close encounters with pericenter distance less than $D$ in one year with perturbers that can cause $e_{\max} \geq 0.1$. Because $N(<D)$ corresponds to the sum of the number of collisions between a target object with a radius of $D$ and objects with a radius ranging from $R_p$ to $R_p + dR_p$, we have

$$N(<D) = \sum_j P_j \left[ \int_{R_{\text{ring}}}^{D} \left(D + R_{p,c}(D)\right)^2 n_j\left(R_{p,c}(D)\right) dD \right], \qquad (10)$$

where, $P_j$ is the intrinsic collision probability between hot classical TNOs and another population $j$ (Abedin et al., 2021), $R_{p,c}(D)$ is the perturber radius equivalent to $M_{p,c}^*(D)$ (Eqs. 4-9), and $n_j(R_p)$ is the differential size frequency distribution (SFD) of the population $j$. We obtain $n_j(R_p)$ from the absolute magnitude distribution expressed by power-law with multiple knees from Abedin et al. (2022), which takes into account the exponentially tapered SFD for cold classical TNOs (Kavelaars et al., 2021) and reflects the observational results from New Horizons spacecraft of Pluto, Charon, and Arrokoth (Singer et al., 2019; Spencer et al., 2020). The absolute magnitude in $r$-band, $H_r$, can be converted to radius of the object, $r'$, with the equation:

$$\log_{10} 2r' = 0.2 m_{\odot,r} + \log_{10} 2r_\oplus - 0.5 \log_{10} p_r - 0.2 H_r, \qquad (11)$$

where $m_{\odot,r}$ ($= -27.07$ mag) is the apparent $r$-band magnitude of the Sun (Gwyn, 2012), $r_\oplus$ is the distance between the Sun and Earth in the same unit used for radius of the object, $r'$. We assume the geometric albedo in $r$-band of $p_r = 0.14$ for cold classical TNOs or $p_r = 0.08$ for other dynamically excited TNOs (Vilenius et al., 2014). We use the density of 1000 kg/m$^3$ to convert $r'$ to $M_p$ (e.g., Parker & Kavelaars, 2012; Nesvorný et al., 2021; Campbell et al., 2023).

We obtain the lifetime of the ring system, $\tau(<D)$, as the inverse of $N(<D)$. The values of $\tau(<D)$ and $N(<D)$ can be obtained for arbitrary $D$. But we calculate with $D_{\max}$ which satisfies $M_{p,c}(D_{\max}) = M_{p,\max}$ and define $\tau_{\text{ring}}$ as $\tau_{\text{ring}} = \tau(<D_{\max}) = 1/N(<D_{\max})$. We assume the largest perturber $M_{p,\max}$ to be five times as massive as Pluto. Note that the size of the largest perturber hardly affects our results because of the dearth of such very large objects. We set the smallest perturber to be the one equivalent to $M^*_{p,c}(R_{\text{ring}})$ (Eqs. 4-9).

Figure 11a shows that $\tau_{\text{ring}} > 10^4$ Gyr for ring systems around Chariklo, Haumea, and Quaoar. Since this is the case of $i_p = 0°$ and $\omega_p = 0°$, $\tau_{\text{ring}}$ should be longer if arbitrary $i_p$ and $\omega_p$ are considered. Therefore, we conclude that disruption due to a close encounter with a small object is not a plausible scenario for ring systems around Chariklo, Haumea, and Quaoar. The putative eccentricity of Chariklo's inner ring is assumed be ~0.004 based on the variation in the ring width (Pan & Wu, 2016). Using the same method, we obtain $\tau_{\text{ring}}$~67 Gyr and 1.9 Gyr in the case of $e_{\max} \geq 10^{-3}$ and $10^{-4}$, respectively. Although $\tau_{\text{ring}}$ in the case of $e_{\max} \geq 10^{-4}$ is less than 4 Gyr, close encounter with a small object may not be a plausible cause of the eccentricity of Chariklo's inner ring because such a change in eccentricity may be quickly damped by dissipative collisions between ring particles.

It should be noted that our calculation ignores the non-axisymmetric gravitational potential, which determines stable regions and induces particle migration (Sicardy et al., 2019; Winter et al., 2019; Marzari, 2020; Sanchez et al., 2020; Sumida et al., 2020; Madeira et al., 2022; Giuliatti Winter et al., 2023; Ribeiro et al., 2023; Rodríguez et al., 2023). Incorporating the effect of the oblateness ($J_2$) and ellipticity ($C_{22}$) of the central body and using the Poincaré surface of section technique, it was demonstrated that particles are stable when the eccentricity is roughly below ~0.2 for Chariklo and Haumea (Giuliatti Winter et al., 2023; Ribeiro et al., 2023). Because our simulation shows that close encounter resulting in $e_{\max}$~0.1 is highly unlikely, particles are expected to remain in the stable region. On the other hand, numerical simulations indicated that particle migration occurs on the timescale of days to months for Chariklo (Giuliatti Winter et al., 2023), and of days (Sumida et al., 2020) to 10 years (Sicardy et al., 2019) for Haumea. However, this long-term stability, for which the non-axisymmetric gravitational potential is essential, is beyond the scope of this study.

**4.2. Dependence on the size of the central body**

Lastly, we investigate the dependence of $\tau_{\text{ring}}$ on $R$, by calculating $\tau_{\text{ring}}$ for $R$ =1000, 500, 250, 100, and 50 km with the same method in Section 4.1. We place a fictitious ring at $3R$ with a width of $0.05R$ based on the ring radius of Chariklo and Haumea (Table 3). For smaller central bodies ($R$ =25, 10, 5, 1, 0.5, 0.1 km), we obtain $M_{p,c}^*(D)$ analytically from ZT04Imp and then $\tau_{\text{ring}}$. This is because $T_e/T_{\text{ring}}$ becomes smaller for smaller central bodies, and the impulse approximation is more suitable for the estimation of $e_{\max}$ and $M_{p,c}^*(D)$.

Figure 11a shows that $\tau_{\text{ring}}$ increases as the central body becomes larger: $\tau_{\text{ring}}$ in the case of $R$ =1000 km is approximately two orders of magnitude longer than that of $R$ =100 km. Even for km-sized central bodies, $\tau_{\text{ring}}$ is longer than 4 Gyr. Therefore, in terms of a close encounter with a small object, larger central bodies are more likely to retain their ring systems but 100 m-sized central bodies can also retain their ring systems. Additionally, the critical mass ratio at ring radius, $M_{p,c}^*(R_{\text{ring}})$, decreases as the central body becomes larger (Fig. 11b). Thus, ring systems around larger central bodies require perturbers with smaller mass ratio in order to cause the same degree of disturbance. With constant $v_{\text{peri}}$, the timescale of close encounter becomes longer for larger central bodies. This gives perturbers longer time to interact with ring particles and thus less massive perturber are needed for comparable disturbance. Note that mass of the perturber, $M_p$, corresponding to $M_{p,c}^*(R_{\text{ring}})$, increases as the central body becomes larger (Fig. 11c). Moreover, the response of shepherding satellites to close encounters should be similar to that of ring particles because they are presumed to be small and orbit in vicinity of the ring. Hence, if a ring is maintained by shepherding satellites, the lifetime of the ring corresponds to $\tau_{\text{ring}}$ shown in Fig. 11a. Therefore, a ring maintained by shepherding satellites would also remain nearly undisturbed by close encounter over the timescale longer than the age of the solar system.

Ring systems are not found to date around near-Earth asteroids (NEAs) and main-belt asteroids (MBAs). The intrinsic collision probability for both NEAs and MBAs is on the order of $10^{-18}$ (e.g., Bottke et al., 1994a, 1994b), which is four or five orders of magnitude larger than that for TNOs (Abedin et al., 2021). In addition, the estimated number of MBAs (Maeda et al., 2021) does not differ significantly with that of TNOs (e.g., Kavelaars et al., 2021; Petit et al., 2023). Thus, $\tau_{\text{ring}}$ of NEAs and MBAs would be on the order of 100 Myr for km-sized asteroids and on the order of 10 Myr for 100 m-sized asteroids, which may explain the absence of ring systems in addition to other processes. For NEAs, however, Fang & Margot (2012) showed that the timescale of close encounter within 20-30 Earth radii is ~1 Myr, considerably shorter than that of close encounter with another NEA. Hence, close encounters with the Earth and possibly Mercury and Venus are more likely to have a greater impact on the lifetime of potential ring systems in the inner solar system.

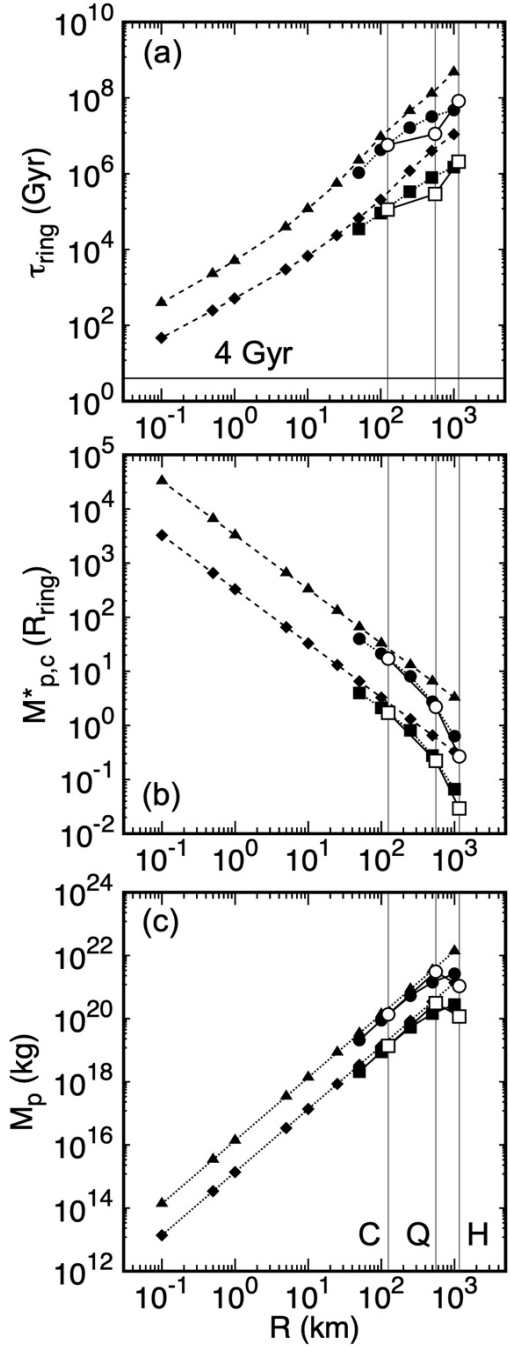

**Fig. 11.** (a) Lifetime of ring systems, $\tau_{\text{ring}}$, as a function of $R$. Open squares and circles represent $\tau_{\text{ring}}$ for $e_{\max} \geq 0.1$ and $1$. The black lines show the size of Chariklo (C), Haumea (H), and Quaoar (Q). Filled squares and circles indicate $\tau_{\text{ring}}$ for $e_{\max} \geq 0.1$ and $1$, respectively. Lastly, $\tau_{\text{ring}}$ for $e_{\max} \geq 0.1$ and $1$ calculated from ZT04Imp is denoted by filled diamonds and triangles, respectively. The black horizontal line shows the age of the solar system, 4 Gyr. (b) Same with (a) but for critical mass ratio at ring radius, $M^*_{p,c}(R_{\text{ring}})$, that can cause $e_{\max} \geq 0.1$ (filled squares) of the fictitious ring systems. Mass ratio that can cause $e_{\max} = 1$ at $D = R_{\text{ring}}$

is also shown (filled circles). (c) Critical mass ratio, $M_{p,c}^*(R_{\text{ring}})$, shown in the panel (b) is converted to mass, $M_p$.

## 5. Conclusion

We found that significant disruption of the ring would not occur unless the perturber is as massive as or more massive than the central body. Even when close encounter occurs very close to the ring, $M_{p,c}^*$ is on the order of $10^{-1}$ for Chariklo's ring and on the order of $10^{-2}$ for Haumea's and Quaoar's rings. Our results show that $e_{\max}$ and $\delta e$ have a linear relationship with $M_p^*$ for ring systems around Chariklo, Haumea, and Quaoar. In the case of Chariklo's and Quaoar' rings, $\delta e$ follows a power law of $D$, and the impulse approximation given by Zakamska & Tremaine (2004) (ZT04Imp) matches well. For Haumea's ring, $\delta e$ displays an exponential relationship with $D$, and the analytical expression called "exponential regime" given by Heggie & Rasio (1996) (HR96Exp) agrees with ZT04Imp. We consider that $\delta e$ follows HR96Exp because $T_{\text{ring}}$ and $T_e$ are comparable for Haumea's ring. For Chariklo's ring and Quaoar's ring, the condition for the impulse approximation ($T_e/T_{\text{ring}} \ll 1$) is satisfied and thus $e_{\max}$ and $\delta e$ are in good accordance with ZT04Imp. Other expressions we considered did not match well because they are based on secular theories which are only applicable when $T_e/T_{\text{ring}} \gg 1$. We also found that depending on the orbital elements, $i_p$ and $\omega_p$, $e_{\max}$ varies approximately one order of magnitude. We note that our calculations ignore the oblateness of the central body or the interparticle collisions. Further studies are necessary to fully understand the fate of the disturbed ring.

We calculated the number of close encounters that can result in $e_{\max} \geq 0.1$ and obtained the lifetime. We found that the lifetimes of the ring systems around Chariklo, Haumea, and Quaoar are much longer than the age of the solar system. Therefore, we conclude that the ring systems around Chariklo, Haumea, and Quaoar are highly unlikely to experience close encounters that bring about a severe damage on the ring. Furthermore, we carried out a similar series of calculations with fictitious ring systems changing the radius of the central body. Our results indicate that ring systems around larger objects require larger perturbers. In terms of mass ratio, however, larger central bodies require smaller $M_{p,c}^*$. The lifetime is longer for ring systems around larger objects. Still, the lifetime of ring systems around 100 m-sized asteroids is longer than the age of the solar system. On the other hand, the lifetime is on the order of 100 Myr for km-sized asteroids and 10 Myr for 100 m-sized asteroids in the inner solar system, which may explain why ring systems have not been found around NEAs or MBAs.

**Acknowledgement**

The authors wish to thank two anonymous reviewers for their comments, which significantly tightened the manuscript. This work was supported by JSPS KAKENHI Grant Number 23KJ1566 for RI and partly supported by the JSPS Grants-in-Aid for Scientific Research (Nos. 20K14538 and 20H04614) for NH. We wish to thank Keiji Ohtsuki for his helpful discussions and inputs to this study.